\newif\ifieee
\newif\ifacm
\acmtrue

\ifieee
\RequirePackage[OT1]{fontenc}
\documentclass[conference,compsoc]{IEEEtran}
\fi
\ifacm 
\documentclass[sigconf, anonymous=False, authorversion, nonacm]{acmart}

\acmYear{2024}\copyrightyear{2024}
\setcopyright{cc}
\acmConference[ASIA CCS '24]{ACM Asia Conference on Computer and Communications Security}{July 1--5, 2024}{Singapore, Singapore}
\acmBooktitle{ACM Asia Conference on Computer and Communications Security (ASIA CCS '24), July 1--5, 2024, Singapore, Singapore}
\acmDOI{10.1145/3634737.3661141}
\acmISBN{979-8-4007-0482-6/24/07}

\keywords{Directed Fuzzing, Software Security, Target Selection, Software Metrics}

\ccsdesc[500]{Security and privacy~Software and application security}
\ccsdesc[500]{Security and privacy~Systems security}
\ccsdesc[100]{General and reference~Surveys and overviews}

\makeatletter
\renewcommand{\@copyrightpermission}{%
  {\footnotesize  
  This is a preprint of the paper published in the proceedings of the 19th ACM Asia Conference on Computer and Communications Security (AsiaCCS '24). The final version is available at \href{https://doi.org/10.1145/3634737.3661141}{https://doi.org/10.1145/3634737.3661141}
  } \\}
\makeatother 

\fi

\usepackage[utf8]{inputenc}
\usepackage[english]{babel}

\usepackage{microtype}

\usepackage{xcolor}

\usepackage{blindtext}

\usepackage{rotating}

\usepackage{booktabs}
\usepackage{multirow}
\usepackage{tabularx}
\newcolumntype{C}[1]{>{\centering\arraybackslash}p{#1}}
\newcolumntype{R}[1]{>{\raggedleft\arraybackslash}p{#1}}
\usepackage{colortbl}

\usepackage{graphicx}
\graphicspath{ {./images/} }

\usepackage{svg}
\svgpath{ {./images/} }

\usepackage{amsmath}
\usepackage{mathtools}

\usepackage{tikz}
\usetikzlibrary{matrix}
\usetikzlibrary{patterns}
\usetikzlibrary{calc}

\usepackage[most]{tcolorbox}

\usepackage{nopageno}

\tcbset{
  base/.style={
    arc=1.5mm, % Adjust the arc parameter for rounded corners
    bottomtitle=0.5mm,
    boxrule=0mm,
    colbacktitle=black!10!white, 
    coltitle=black, 
    left=2.5mm,
    leftrule=1mm,
    right=3.5mm,
    title={#1},
    toptitle=1.2mm, 
  }
}

\definecolor{brandblue}{rgb}{0.34, 0.7, 1}
\newtcolorbox{mainbox}[1]{
  colframe=brandblue, 
  base={#1}
}

\newtcolorbox{subbox}[1]{
  colframe=black!30!white,
  base={#1}
}

\usepackage{pgfplots}
\usepgfplotslibrary{groupplots}
\pgfplotsset{compat=1.18}

\usepackage[position=top]{subfig}

\usepackage{amsmath}

\usepackage{pifont} % used to get \ding command
\usepackage{wasysym} % more symbols, like circle, rightcircle,...

\usepackage{bbding}

\ifieee
\usepackage[numbers]{natbib}
\fi
\ifacm
\usepackage{natbib}
\fi

\definecolor{coldefault}{RGB}{11, 49, 66}

\definecolor{colpositive}{RGB}{11, 49, 66}
\definecolor{colnegative}{RGB}{170,72,72}

\definecolor{colwarning}{RGB}{219, 172, 20}
\definecolor{colrelevant}{RGB}{222, 96, 98}
\definecolor{colirrelevant}{RGB}{222, 96, 98}

\definecolor{collinevul}{RGB}{252, 90, 40}
\definecolor{colreveal}{RGB}{252, 90, 40}
\definecolor{colcodet5p}{RGB}{252, 90, 40}
\definecolor{colrats}{RGB}{152, 190, 87}
\definecolor{colcppcheck}{RGB}{152, 190, 87}
\definecolor{colcomplexity}{RGB}{11, 49, 66}
\definecolor{colvulnerability}{RGB}{11, 49, 66}
\definecolor{colsan}{RGB}{186, 157, 157}
\definecolor{colasan}{RGB}{186, 157, 157}
\definecolor{colubsan}{RGB}{186, 157, 157}
\definecolor{colmsan}{RGB}{186, 157, 157}
\definecolor{colrecent}{RGB}{90, 132, 164}
\definecolor{colrandom}{RGB}{0, 0, 0}

\newcommand{\change}[1]{\textcolor{black}{#1}}
\newcommand{\cstart}{\color{black}}
\newcommand{\cend}{\color{black}}

\newcommand{\relevancescore}{r}
\newcommand{\functionlocs}{\mathbf{F}}
\newcommand{\functionlocssubset}{F}

\newcommand{\functionloc}{\mathsf{f}}
\newcommand{\relevanceoracle}{\mathrm{O}}

\newcommand{\ossfuzz}{OSS-Fuzz}
\newcommand{\positive}[1]{\textcolor{colpositive}{#1}}
\newcommand{\negative}[1]{\textcolor{colnegative}{#1}}

\newcommand{\boldpar}[1]{\medskip\noindent\textbf{#1}.\xspace}
\newcommand{\italicpar}[1]{\medskip\noindent\emph{#1}.\xspace}
\newcommand{\cmark}{\positive{\ding{51}}}
\newcommand{\xmark}{\negative{\ding{55}}}

\newcommand{\stepone}{\ding{182}\xspace}
\newcommand{\steptwo}{\ding{183}\xspace}
\newcommand{\stepthree}{\ding{184}\xspace}
\newcommand{\stepfour}{\ding{185}\xspace}

\newcommand{\numanalyzedpapers}{25\xspace}

\title[Assessing Target Selection Methods in Directed Fuzzing]{SoK: Where to Fuzz? \\ Assessing Target Selection Methods in Directed Fuzzing}

\ifacm \author{Felix Weissberg}
\affiliation{
    \institution{TU Berlin \& BIFOLD}
    \country{Germany}
}
\author{Jonas Möller}
\affiliation{
    \institution{TU Berlin \& BIFOLD}
    \country{Germany}
}
\author{Tom Ganz}
\affiliation{
    \institution{SAP Security Research}
    \country{Germany}
}
\author{Erik Imgrund}
\affiliation{
    \institution{SAP Security Research}
    \country{Germany}
}
\author{Lukas Pirch}
\affiliation{
    \institution{TU Berlin \& BIFOLD}
    \country{Germany}
}
\author{Lukas Seidel}
\affiliation{
    \institution{Binarly}
    \country{USA}
}
\author{Moritz Schloegel}
\affiliation{
    \institution{CISPA}
    \country{Germany}
}
\author{Thorsten Eisenhofer}
\affiliation{
    \institution{TU Berlin \& BIFOLD}
    \country{Germany}
}
\author{Konrad Rieck}
\affiliation{
    \institution{TU Berlin, BIFOLD \& TU Wien}
    \country{Germany}
}

 \fi

\begin{document}

% Authors (IEEE)
\ifieee \author{Anonymous Submission} \fi

% Title (IEEE)
\ifieee \maketitle \fi

% Abstract
\begin{abstract}
	A common paradigm for improving fuzzing performance is to focus on selected regions of a program rather than its entirety. While previous work has largely explored \emph{how} these locations can be reached, their selection, that is, the \emph{where}, has received little attention so far. 
In this paper, we fill this gap and present the first comprehensive analysis of target selection methods for fuzzing. To this end, we examine papers from leading security and software engineering conferences, identifying prevalent methods for choosing targets. By modeling these methods as general scoring functions, we are able to compare and measure their efficacy on a corpus of more than 1,600 crashes from the \ossfuzz{} project. 
Our analysis provides new insights for target selection in practice: 
% %
\change{
First, we find that simple software metrics significantly outperform other methods, including common heuristics used in directed fuzzing, such as recently modified code or locations with sanitizer instrumentation. Next to this, we identify language models as a promising choice for target selection. In summary, our work offers a new perspective on directed fuzzing, emphasizing the role of target selection as an orthogonal dimension to improve performance.
}
\end{abstract}

% Title
\ifacm \maketitle \fi

% Keywords (IEEE)
\ifieee
%\begin{IEEEkeywords}
%component, formatting, style, styling, insert
%\end{IEEEkeywords}
\fi

% Sections
\section{Introduction}
\label{sec:introduction}
\begin{figure*}[t!]
	\centering
	\includegraphics[width=0.8\linewidth]{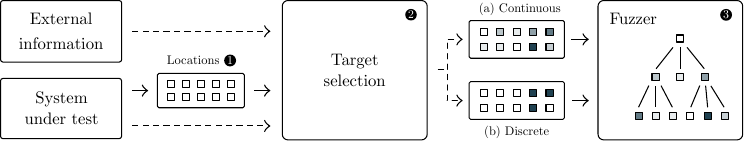}
	\vspace*{1em}
	\caption{\textbf{Target selection.} \normalfont The initial step in target selection is the extraction of code locations from the SUT (Step~\stepone). The granularity of this extraction depends on the specifics of the selection method and could, for example, be on a function-level or basic-block level. After extraction, the locations and (optionally) the SUT or external information (e.g., code change timestamps) are forwarded to the target selection method. This method then assigns a score to each location (Step~\steptwo), which is either (a) continuous or (b) discrete. Finally, the fuzzer utilizes the annotated code locations for guidance (Step~\stepthree).}
	\label{fig:target-selection}
\end{figure*}

Fuzzing has been a thriving area of security research in recent years. Numerous techniques have been proposed to improve the efficacy of fuzzers, ranging from concepts to increase efficiency~\citep{CheChe18, AscSchBlaGaw+19, ShePeiEpsYan+19, SeiMaiMue23} and mitigate roadblocks~\citep{BlaAscSchAbb+19, BarSchSchSch+23} to application-specific testing strategies~\citep{CheDiaZhaZuo+18, SchChlSchBar+23, RawJaiKumCoj+17}. This research encompasses a large variety of technical contributions inspired from software engineering, compiler design, and software security, yet one general paradigm stands out as a key concept in many of the approaches to improving fuzzing performance: \emph{target selection}.

Rather than treating all code locations equally, many approaches guide the fuzzer towards areas more likely to contain defects, such as those involving memory accesses, insecure API calls, or recent patches~\citep[]{OstRazBosGiu+20,HalSloNeuBos+13, MarCad+13}. Similarly, several fuzzers direct the testing explicitly to a chosen set of locations to improve efficiency in different stages of software development~\cite[][]{BöhPhaNguRoy+17, CheXueLiChe+18, HuaGuoShiYao+22}. Although these approaches fundamentally differ in \emph{how} they reach these regions and guide the fuzzer, they all share the concept of focusing on selected target locations rather than uniformly covering the program as traditional fuzzers~\citep{SutGreAmi07}.

Surprisingly, the analysis of suitable targets for a fuzzer, that is, the \emph{where} of the paradigm, has received little attention so far. While some fuzzers incorporate a fixed strategy for locating targets, such as focusing on sanitizers~\citep{OstRazBosGiu+20}, code changes~\citep{MarCad+13}, or buffer operations~\citep{HalSloNeuBos+13}, several approaches for directed fuzzing leave the target selection to the practitioner. As a result, it is currently unclear which selection method performs best in practice, and the simple question---\emph{where to fuzz?}---is still unexplored.

In this paper, we aim to bridge this gap and present the first comprehensive analysis of \emph{target selection methods} for directed fuzzing. To this end, we conduct a literature review of \numanalyzedpapers papers concerned with directed fuzzing published at top-tier security and software engineering conferences over the past five years. We identify two dimensions that characterize the selection of targets: (1)~the information source and (2)~the selection mechanism. For example, some selection methods are based on analyzing binary code, while others require the source code or further external information. Similarly, we can categorize the selection mechanisms into metric-based and pattern-based heuristics, identifying interesting code locations for the fuzzer to inspect.

Based on this systematization, we proceed to compare different methods for target selection. To avoid evaluating the efficacy of fuzzers rather than their targets, we conduct this comparison in isolation. That is, we model target selection methods as abstract \emph{scoring functions} that retrieve a set of code units and assign scores to them, indicating their ``interestingness'' for the particular fuzzing setup. This approach is applicable to any selection method and differs only in (1) which code units are provided and (2)~how they are ranked according to the underlying scoring functions. As a result, we are able to formulate our evaluation as an information retrieval task and assess the methods ``in~vitro'' using corresponding performance measures.

For our evaluation, we assemble a dataset of more than 1,600 crashes from 97 software projects of \ossfuzz{}~\cite{Ser+23}, which to the best of our knowledge represents the largest corpus of reproducible crashes to date. These crashes serve as ground truth and allow for quantitatively measuring how well a target selection method and the underlying scoring function can retrieve code that actually triggers crashes during a fuzzing campaign. By measuring this retrieval performance in different settings, we can draw general conclusions about target locations and derive recommendations for improving directed fuzzing, independent of specific testing strategies.

Our evaluation uncovers different insights for target selection: First, we observe that simple software metrics, such as vulnerability scores of the Leopard framework~\cite{DuCheLiGuo+19}, significantly outperform all other approaches. This holds true across all types of crashes and indicates that advanced methods for target selection are not yet able to surpass simple numerical metrics. Second, we identify large language models for code, such as CodeT5+, as promising alternatives that almost reach the efficacy of software metrics.

In summary, our work opens a new perspective on directed fuzzing by decoupling the fuzzing mechanism (how) from the selection of target locations (where). We are the first to observe that selection methods differ significantly when studied in isolation, and that some strategies, such as software metrics, consistently outperform other methods. Given the simple nature of these metrics, we argue that developing better selection methods is a promising path for future research and that machine learning models can potentially serve as a means to success in this area.

\boldpar{Contributions} In summary, we make the following
major contributions in this work:

\begin{enumerate}
\setlength{\itemsep}{4pt}

\item \emph{Systematization of target selection.} We conduct the first systematic analysis of target selection methods for directed fuzzing. For this, we analyze a total of \numanalyzedpapers papers related to target selection published between 2018--2023 at top security and software engineering conferences.

\item \emph{Large-scale evaluation.} We model the task of target selection as an information retrieval problem and assemble a dataset of more than 1,600 reproducible crashes as ground truth. This is the largest available dataset of this kind, and we provide it publicly as part of this paper's artifacts.

\item \emph{Insights for target selection in fuzzing.}  Based on the results of our analysis, we provide insights and recommendations for target selection in different fuzzing contexts.
\end{enumerate}

\section{Fuzzing Specific Code Locations}
\label{sec:background}
The ultimate goal of fuzzing is to uncover bugs. When testing unknown software, the underlying defect distribution, i.e., the number and location of defects, is naturally unknown. In this case, code coverage has proven an excellent proxy metric: after all, we cannot find flaws in code that has not been executed by the fuzzer. In fact, recent research has found a strong correlation between code coverage and software defects~\cite{BöhSzeMet22}. However, the fuzzer performing best in terms of coverage may not necessarily be the one finding the most bugs. Still, using code coverage as \emph{feedback} for guidance has become the de-facto standard of modern fuzzing, offering excellent exploration capabilities in absence of knowledge of the underlying defect distribution.

In certain scenarios, however, we may know, or at least can assume, the (potential) location of defects. For example, new and untested code is more likely to contain bugs than well-tested software~\cite{ZhuBöh21}. Similarly, a static analysis tool can pinpoint specific code locations as potentially buggy. In such cases, \emph{guiding} the fuzzer towards these locations may help to uncover defects faster than when indifferently exploring all parts of the program. This insight lead to the introduction of \emph{directed fuzzers}: Traditionally, these fuzzers accept one or more code locations, for example, addresses in the binary or source code lines, that a fuzzer should target during execution. While some fuzzers directly turn their focus towards these targets, others initially explore the program similarly to regular fuzzers before switching to an exploitation phase where they focus on the desired target locations~\cite{BöhPhaNguRoy+17}. 

In this work, we consider a wider definition of directed fuzzing: We not only include traditional approaches that accept predefined target locations but also consider any approach that uses some metric (beyond coverage feedback) to guide the fuzzer to specific locations. A prime example of such an approach is \emph{regression greybox fuzzing}~\cite{ZhuBöh21}, where testing is directed towards recently changed code rather than user-specified targets. While different to traditional directed fuzzing that uses a fixed set of targets, this technique still \emph{directs} the fuzzer to specific locations. The difference merely lies in the \emph{target selection} and the type of \emph{scoring} used to measure the ``relevance'' of a given code location. In other words, the score of code locations tells the fuzzer whether to prioritize them during fuzzing. 
Figure~\ref{fig:target-selection} outlines the overall process enabling the guidance of the fuzzer. \change{Before discussing the individual steps and focusing on the \emph{target selection}, we first motivate its relevance towards improving the probability of finding a crash. Following this, we examine the two different types of scoring used in directed fuzzing in more detail.}

\cstart
\boldpar{Relevance of target selection}
When examining prior works that present directed fuzzers, we can make two observations. First, these studies typically demonstrate---as part of their experiments---that their directed fuzzing approach outperforms an undirected coverage-guided fuzzer that serves as a baseline~\cite{BoePhaNguRoy17, CheXueLiChe+18, WüsChr20, NguBarBonGro+20, AscSchAbbHol+20, ZhaLiuYaoJin+23}. Second, they show their fuzzers are in fact directed; that is, they gradually steer the fuzzing process towards the provided targets~\cite{BoePhaNguRoy17, CheXueLiChe+18, WüsChr20}. 
If we assume all targets are unrelated to crash sites, a directed fuzzer will spend resources to focus on code that contains no bugs, meaning the directed fuzzer is unlikely to outperform a coverage-guided one in this case.
Thus, we conclude that selecting relevant targets is indeed a crucial step in directed fuzzing to increase the probability of finding crashes.
In other words, not only implementation details such as fuzzing throughput determine the fuzzer's success but also the target selection. Despite its relevance to the success of directed fuzzing, we find this aspect is often overlooked and not considered individually.
With this in mind, we now take a closer look at the two types of scoring mechanisms that enable a selection of targets.

\cend

\boldpar{Discrete scoring}
The first type of scoring assigns a \emph{discrete} value, usually either 0 or 1, to each location. Hence, it differentiates between relevant locations and irrelevant ones. This scoring type closely resembles most tools traditionally referred to as directed fuzzers: They accept a set of target locations, marking them as relevant, whereas every other code location is considered irrelevant. The target selection decides upon the metric according to which target locations are chosen. During execution, the \emph{distance} to relevant code locations is then used as a proxy score that allows to indirectly rank and compare locations.

\boldpar{Continuous scoring}
The second type of target selection assigns a \emph{continuous} value to all code locations: This score can, for example, be based on the last time this location has been modified~\citep{ZhuBöh21}, the number of sanitizer primitives it contains~\citep{OstRazBosGiu+20}, or a software metric describing its code complexity~\citep{DuCheLiGuo+19}. As each code location is assigned its own score, these approaches usually do not need a proxy metric, such as the distance to a target location, and the fuzzer can iteratively optimize over the scores of the visited code regions during its operation.

\boldpar{Comparison of scoring functions}
The two types of scoring functions lead to a different effectiveness of target selection. To examine this difference, let us consider a program that contains a single defect. With continuous scoring, the probability that this bug is triggered increases steadily with the quality of the scores, as a high quality scoring means that the defective locations are assigned significantly higher scores than non-defective ones.

In contrast, when we consider a discrete scoring, we usually focus on a small number of targets. In this case, a low-quality scoring may not flag the error location as relevant, and hence it may become unlikely that the fuzzer reaches the bug at all, as it is effectively steered towards non-defective locations. However, once the location is marked, our chances increase noticeably.
When evaluating target selection methods in Section~\ref{sec:results}, we take this difference into account.

In comparison, discrete scoring provides more versatility: Regardless of how the target set was derived, guiding a fuzzer using the distance to its locations will work, providing greater versatility compared to scoring each code locating using a specific metric. Changing the metric according to which targets are selected is supported by design: Often, these tools allow the user to specify arbitrary code locations. 
On the other side, all selected locations are treated equally by discrete scoring functions. If two of these targets are not equally relevant, we lose information, as individual locations cannot be ranked differently.
In contrast, continuous scores do not suffer from this problem and can differ between targets on a more fine-granular basis.

\medskip
In short, we can distinguish target selection approaches based on the score they assign to their targets, with each type having advantages and disadvantages.

\boldpar{Overview}
With the two types of scoring in mind, we can focus on their place in the overall fuzzing process. Figure~\ref{fig:target-selection} provides an overview of the general flow: Based on external information known a priori or information extracted from the System Under Test (SUT), we know \emph{what} code to target. Our target selection receives this metric and the SUT (Step \stepone). Based on the available information, the \emph{target selection} (Step \steptwo) can be performed by assigning a \emph{discrete} or \emph{continuous} score to each code location. Finally, the fuzzer tests the SUT (Step \stepthree) while focusing on the targets.

\section{Systematization of Target Selection}
\label{sec:localization}
Selecting good targets is crucial to the success of directed fuzzing. Yet, we find that this aspect has surprisingly received little attention in fuzzing research so far. Consequently, before turning towards a comparative analysis of different strategies for locating interesting code, we first conduct a comprehensive literature review on target selection methods used currently for directed fuzzing.

\italicpar{Literature} For our review, 
we investigate papers published at top security and software engineering venues between 2018--2023. In particular, we collect all papers from the following A$^*$ conferences~\cite{coreranking}: ASE, FSE, ICSE, CCS, NDSS, USENIX Security, and S\&P. After filtering out papers not related to fuzzing, this leaves us with 289 fuzzing papers. We then automatically identify papers focusing on \emph{directed fuzzing} by filtering out any paper that does not contain the word ``directed'' at least three times. This heuristic is grounded in the assumption that publications in a field are likely to either name the field multiple times or at least mention it when comparing against prior works. Using this process, we end up with 31 publications. Manually analyzing all of them yields \numanalyzedpapers papers that deal with directed fuzzing under our definition. 

\italicpar{Review method}
\change{To distinguish between different target selection techniques, we review all papers in-depth and distill the working principle for locating targets. As result of this process, we arrive at four characteristics to categorize selection methods. First, we consider the source from which the information originates that is used to direct the fuzzer. We refer to this as the \emph{information source}. Second, we distinguish whether the underlying scoring function is \emph{discrete} or \emph{continuous} as discussed previously.
We label this as the \emph{scoring type} of the selection method. Third, we distinguish between different levels of granularity of the target selection method. We denote this as the \emph{granularity}. Lastly, we differentiate target selection techniques with respect to what their scoring mechanism is based on. We refer to this property as the target selection's \emph{scoring mechanism}. 
}

\italicpar{Analysis results}
The results of our literature review are shown in Table~\ref{tab:survey}. For each publication, we indicate whether a paper put a focus on the target selection (denoted as \cmark{}) or if it exclusively used targets obtained from third parties, such as previous literature or stack traces of crashes from a bug tracker (denoted as \xmark{}). We further record whether the employed target selection was continuous (C) or discrete (D). 
In the following, we shift our focus to the information sources, scoring mechanism and granularity used within the examined publications.

\boldpar{The scoring mechanism landscape} 
From our literature review, we observe two broad categories of scoring mechanisms: one based on metrics and one on patterns. The metrics-based category includes methods that utilize code metrics to score individual code locations. In contrast, the pattern-based target selection category utilizes heuristics for this purpose.

\begin{table}[t]
    \centering
    \footnotesize
    \caption{\change{\textbf{Overview of target selection methods.} \normalfont 
    The \emph{Where?} column denotes if a publication puts a focus on the target selection method (\cmark) or if it exclusively relied on a target selection from previous work or reproduction tasks (\xmark). Column \textit{Scoring type} records if a discrete target selection methods (D) or a continuous one (C) is used.
    The granularity of the target selection method is categorized in \textit{Granularity}: instructions (I), basic blocks (B), statements (S), lines (L), and functions (F). The \textit{Source} and the \textit{Scoring} column provide further details about the origin of the additional information and utilized core method of the target selection, respectively.}}
    % \vspace{-0.1cm}
		\cstart
\resizebox{\linewidth}{!}{
\begin{tabularx}{0.88\linewidth}{@{}
                             p{1.9cm} % Paper
                             R{0.9cm} % Venue
                             C{0.175cm} % Where?
                             C{0.175cm} % Scoring type
                             C{0.175cm} % Granularity
                             C{0.175cm}C{0.175cm}C{0.175cm} % Source
							 C{0.175cm}C{0.175cm}} % Scoring
\toprule 
	& & & & &
  \multicolumn{3}{c}{Source} &
  \multicolumn{2}{c}{Scoring}
  \\
 \cmidrule(lr){6-8} \cmidrule(lr){9-10}
  Paper &
  Venue &
  \rotatebox{90}{Where?}&
  \rotatebox{90}{Scoring type} &
  \rotatebox{90}{Granularity} &
  \rotatebox{90}{Source code} & 
  \rotatebox{90}{Binary code} & 
  \rotatebox{90}{External info} &
  \rotatebox{90}{Metrics-based} &
  \rotatebox{90}{Pattern-based} \\
  \midrule
	SelectFuzz \cite{LuoMenLi23}          & SP'23         &  \xmark & D & L & \Circle & \Circle & \CIRCLE & \Circle & \CIRCLE \\
	ODDFuzz \cite{CaoHeSunOuy+23}      & SP'23         & \cmark & D & F & \CIRCLE & \Circle & \Circle & \Circle & \CIRCLE \\
	StrawFuzzer \cite{ZhaLiaXiaZha+22}     & SP'22         & \cmark & D & F & \CIRCLE & \Circle & \Circle & \Circle & \CIRCLE \\
	GREBE \cite{LinCheWuMu+22}       & SP'22       & \xmark & D & S & \Circle & \Circle & \CIRCLE & \Circle & \CIRCLE \\
	BEACON \cite{HuaGuoShiYao+22}     & SP'22         & \xmark & D & L & \Circle & \Circle & \CIRCLE & \Circle & \CIRCLE \\
	She et al. \cite{SheShaJan22}         & SP'22       & \cmark & C & S & \CIRCLE & \Circle & \Circle & \CIRCLE & \Circle \\
	SAVIOR \cite{CheLiXuGuo+20}       & SP'20         & \cmark & D & B & \Circle & \LEFTcircle & \Circle & \Circle & \CIRCLE \\ 
	CollAFL \cite{GanZhaQinTu+18}      & SP'18         & \cmark & C & B & \Circle & \CIRCLE & \Circle & \CIRCLE & \Circle \\
	DAFL \cite{KimChoHeoCha+23}     & SEC'23     & \xmark & D & L & \Circle & \Circle & \CIRCLE & \Circle & \CIRCLE \\
	DDRace \cite{YuaZhaLiLia+23}      & SEC'23     & \hspace*{-0.11cm} (\cmark) & D & I & \Circle & \LEFTcircle & \Circle & \Circle & \CIRCLE \\
	FishFuzz \cite{ZheZhaHuaRen+23}     & SEC'23     & \cmark & D & F & \Circle & \CIRCLE & \Circle & \Circle & \CIRCLE \\
	AmpFuzz \cite{KruGriRos22}         & SEC'22       & \cmark & D & F & \CIRCLE & \Circle & \Circle & \Circle & \CIRCLE \\
	Lee et al. \cite{LeeShiLee21}         & SEC'21     & \xmark & D & L & \Circle & \Circle & \CIRCLE & \Circle & \CIRCLE \\
	FuzzGuard \cite{ZonLvWanDen+20}      & SEC'20     & \xmark & D & L & \Circle & \Circle & \CIRCLE & \Circle & \CIRCLE \\
	ParmeSan \cite{OstRazBosGiu+20}     & SEC'20     & \cmark & D & B & \Circle & \CIRCLE & \Circle & \Circle & \CIRCLE \\ 
	Jiang et al. \cite{JiaBaiLuHu+22}       & NDSS'22       & \cmark & D & F & \Circle & \Circle & \CIRCLE & \Circle & \CIRCLE \\
	TortoiseFuzz \cite{WanJiaLiuZen+20}     & NDSS'20       & \cmark & C & B & \Circle & \CIRCLE & \CIRCLE & \CIRCLE & \Circle \\
	MC$^2$ \cite{ShaSheSadSin+22}     & CCS'22         & \xmark & D & L & \Circle & \CIRCLE & \CIRCLE & \Circle & \CIRCLE \\
	AFLChurn \cite{ZhuBöh21}						 & CCS'21        & \cmark & C & L & \Circle & \Circle & \CIRCLE & \Circle & \CIRCLE \\
	VulScope \cite{DaiZhaXuLyu+21}       & CCS'21       & \xmark & D & F & \Circle & \Circle & \CIRCLE & \Circle & \CIRCLE \\
	Hawkeye \cite{CheXueLiChe+18}      & CCS'18       & \xmark & D & L & \Circle & \Circle & \CIRCLE & \Circle & \CIRCLE \\
	WindRanger \cite{DuLiLiuMao+22}       & ICSE'22       & \xmark & D & L & \Circle & \Circle & \CIRCLE & \Circle & \CIRCLE \\
	$\text{AFL}_\text{LTL}$ \cite{MenDonLiBes+22}         & ICSE'22       & \cmark & D & S & \Circle & \Circle & \CIRCLE & \Circle & \CIRCLE \\
	Wüstholz et al. \cite{WüsChr20} & ICSE'20       & \xmark & D & L & \Circle & \Circle & \Circle & \Circle & \Circle \\
	Leopard \cite{DuCheLiGuo+19}       & ICSE'19       & \cmark & C & F & \CIRCLE & \Circle & \Circle & \CIRCLE & \Circle \\
\bottomrule
\end{tabularx}}
\cend

    \label{tab:survey}
\end{table}

Out of the 25 directed fuzzing approaches, 14 put a focus on the target selection method (\cmark in the \emph{Where?} column of Table~\ref{tab:survey}). From these 14, ten are pattern-based and use heuristics to select interesting target locations. For example, ODDFuzz~\cite{CaoHeSunOuy+23} selects deserialization methods in Java as targets for a directed fuzzing approach, StrawFuzzer~\cite{ZhaLiaXiaZha+22} uses data storing instructions as targets to increase the memory footprint of Android services and cause the system to crash, and AmpFuzz~\cite{KruGriRos22} targets network-related functions as part of their approach to find amplification vectors for DDoS attacks. All of these heuristics have in common that they focus on finding a specific type of erroneous behavior. Another heuristic employed by three of the approaches, ParmeSan~\cite{OstRazBosGiu+20}, SAVIOR~\cite{CheLiXuGuo+20}, and FishFuzz~\cite{ZheZhaHuaRen+23}, covers a broader range of potential bugs: Their approach to the target selection is based on the idea that sanitizer instrumentation can function as an indicator for the relevance of a code location to a directed fuzzer. After all, sanitizer instrumentation is added at locations where a bug might occur.

% Code Metrics
Instead of heuristics, five approaches use a metrics-based selection method. In particular, TortoiseFuzz~\cite{WanJiaLiuZen+20} and CollAFL~\cite{GanZhaQinTu+18} focus on the number of memory accesses in their target selection method. In addition, TortoiseFuzz augments this information with the number of security related functions, which they identified by crawling the pages referenced from CVE descriptions. Both the target selection used by She~et~al.~\cite{SheShaJan22} and the one presented by Leopard~\cite{WüsChr20} are based on code metrics. The former uses a graph-based metric that assigns scores to code locations based on how many other uncovered code regions could potentially be reached from it. The latter uses two code metrics, a structural complexity metric and what the authors refer to as ``vulnerability metrics'', which revolve around properties of a code region, such as the number of pointer arithmetic operations or the number of nested control structures.

\boldpar{The information source landscape}
The sources providing input to the target selection methods can be broken down into three origins: source code, binary code, and external information. 

The results of our literature survey show that each information source is used by metrics-based as well as pattern-based methods. However, the specific information source is, unsurprisingly, determined by the individual method. Certain information is only available at the source code level, while others are accessible only on the binary code level. For instance, \emph{sanitizer instrumentation} is not present at the source code level, such that the approaches we examined either resorted to the binary code level or LLVM IR for this purpose. We denote the latter with half-filled circles (\LEFTcircle) in Table~\ref{tab:survey}. The same holds true for \emph{memory access instructions}~\cite{GanZhaQinTu+18, YuaZhaLiLia+23}. On the other hand, the code metrics used by Leopard~\cite{DuCheLiGuo+19} require source code as input. In contrast, AFLChurn~\cite{ZhuBöh21} makes use of the information when a code location was last changed, and TortoiseFuzz~\cite{WanJiaLiuZen+20} uses information crawled from pages referenced by the CVE database. In both cases, the information is neither present in the source code nor in the compiled binary but acquired from external sources.

\cstart
\boldpar{The granularity}
Unlike other systematic literature reviews which focus on the fuzzing approaches themselves~\cite{WanZhoYueLin+24, LiaPeiJiaShe+18}, we characterize the methods used for selecting targets. 
Consequently, in our setting, the granularity refers to the target selection method rather than the characteristics of the proposed directed fuzzer. In other words, while certain fuzzers, such as StrawFuzzer~\cite{ZhaLiaXiaZha+22} or Hawkeye~\cite{CheXueLiChe+18} operate on a basic block granularity, the employed target selection methods provide targets with a function respectively source code line granularity.

In our analysis, we categorize all target selection methods into five distinct levels of granularity: instructions (I), basic blocks (B), statements (S), lines (L), and functions (F). While instruction and basic block granularity only concern methods using the binary code as information source, function granularity can apply to both binary and source code. Line and statement granularity refer to source code lines and individual statements within a line, respectively.
We find that most papers that introduce a new target selection method use a function level granularity. On the other hand, papers that do not introduce a new target selection method are dominated by line level selection methods. This mostly stems from the fact that these papers resort to the source file and line number obtained from tracebacks of crashes as the targets to evaluate their fuzzer.    
\cend

\section{Analysis Framework}
\label{sec:evaluation}
Although numerous target selection mechanisms have been proposed, they rarely receive special attention during the evaluation of the directed fuzzers for which they were designed. Instead, fuzzer evaluations commonly compare different tools against each other so that observed differences cannot confidently be attributed to individual components such as the target selection. 
Furthermore, many evaluations (re-)use the same set of targets (e.g., the dataset AFLGo~\cite{BöhPhaNguRoy+17} presented), thereby measuring performance differences between the newly proposed technique and existing tools, without scrutinizing the strategy for selecting these targets in the first place nor reflecting the various bug classes or target SUTs a selection should perform well on.

\boldpar{Requirements} Therefore, we identify two requirements for conducting a thorough and systematic comparison of target selection mechanisms.
\begin{itemize}
    \setlength{\itemsep}{4pt}
    \item[\textbf{R1.}] We need a suitable method by which we can compare different target selection techniques in isolation, independent of a specific fuzzer ($\rightarrow$~Section~\ref{sec:methodology}).
    \item[\textbf{R2.}] For comparative purposes, a comprehensive ground truth is necessary, ideally representing many different bug classes across a large variety of real-world code
    ($\rightarrow$~Section~\ref{sec:dataset}).
\end{itemize}

\subsection{Target Selection as Information Retrieval}
\label{sec:methodology}
\begin{figure}[t]
	\centering
	\includegraphics[width=0.75\linewidth]{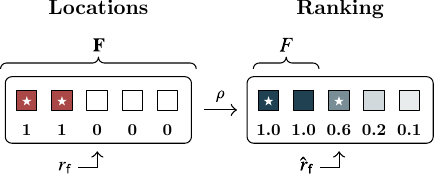}
    \vspace{-0.25em}
	\caption{\textbf{Overview of a retrieval.} \normalfont We compute a ranking using target selection method $\rho$ which assigns each function $\functionloc \in \functionlocs$ a relevance score $\hat{r}_\functionloc$. To measure the quality of target selection method, we compute the $NDCG_k$ for a retrieval $\functionlocssubset$ with cardinality $k$. As ground truth, we use the oracle $\relevanceoracle$ to assign relevance scores $r_\functionloc$ to each location $\functionloc \in \functionlocs$.}
	\label{fig:recallatk}
\end{figure}

As a first step for addressing these requirements,  we present our method for assessing the effectiveness of target selection methods. Our approach is based on the observation that target selection fundamentally resembles an \emph{Information Retrieval} (IR) problem: that is, target selection can be conceptualized as the \emph{retrieval} of relevant target locations in a software project using a scoring mechanism.
In this formulation, targets are \emph{relevant} with regard to the directed fuzzing process, if they benefit the effectiveness of a fuzzer in terms of uncovering new bugs. A fuzzer can use different detection mechanisms to identify whether it triggered a bug; the most common one used by virtually all fuzzers are program crashes, which we focus on subsequently. In other words, targets are relevant if they can point to a code location at which the fuzzer finds a program crash. Similarly, a target can be considered \emph{irrelevant} if it does not contribute to the discovery of a defect and the fuzzer essentially wastes time exploring it. Consequently, a selection methods performs better if it assigns higher scores to relevant code during fuzzing.

Considering target selection as an information retrieval problem yields three key advantages: First, it allows the evaluation of selection methods agnostic to a particular implementation of a directed fuzzer.
Second, it enables to capture different requirements for discrete and continuous target selection methods in form of discrete and continuous retrieval.
Third, we can employ standard evaluation measures from the information retrieval domain to compare and assess the selection on a well-established ground.

In this work, we focus on a function-level granularity as the input for the target selection method.
Intuitively speaking, the optimal target selection method would, thus, precisely \emph{retrieve} the subset $\functionlocssubset$ of all functions $\functionlocs$ within a given software project that lead to a crash. %As the fuzzer must be able to observe this bug, we 
However, given that the exact size of subset $\functionlocssubset$ is typically unknown, we instead consider the $k$ highest-rated functions as determined by the target selection's scoring method.
This scoring can be described as a mapping $\rho$ from functions $\functionloc$ to relevance scores $r$:
$$
   \rho \colon \functionlocs \longrightarrow \mathbb{R}^+, \quad 
   \functionloc \mapsto \relevancescore \,.
$$
For simplicity, we assume that any external information required for the scoring is embedded into the input to $\rho$.
Based on these scores, we then compute a ranking over $\functionlocs$ and retrieve the subset $\functionlocssubset$ as the first $k$ entries in this ranking. Subsequently, subset $\functionlocssubset$ is returned as the output of the target selection.

\boldpar{Ranked retrieval measure}
As common in information retrieval, we distinguish between relevant and irrelevant objects, i.e., whether a bug is present or not.
Thus, to assess the effectiveness of a particular retrieval $\functionlocssubset$, two requirements need to be taken into account: (1) the number of relevant functions in the retrieval, and (2), their respective positions in the ranking.
Metrics like $precision@k$ or $recall@k$ are often used for the former, but fall short in capturing the quality of the ranking. While this would not be a problem for discrete scoring functions, we would lose information for continuous ones. Therefore, we consider the \emph{normalized discounted cumulative gain} (NDCG), a standard performance measure from the information retrieval literature, which accounts for both the relevance of a retrieved function and its rank.

To assign a relevance $\hat{r}_\functionloc$ to each function $\functionloc \in \functionlocs$, we assume there exists an oracle $\relevanceoracle: \functionlocs \rightarrow \{0, 1\}$, that assigns 1 to functions which appeared in the stack trace of a crash, and 0 otherwise.
We can then use this oracle to assign each function a ground truth relevance score $\hat{r}_\functionloc = \relevanceoracle(\functionloc)$.
Functions which are likely to appear in the stack trace of every crash, such as the main function or functions introduced by sanitizers, are assigned a relevance 0, regardless.

Based on these scores and retrieval $F$ with length $k$, we compute the $\text{NDCG}_k$ in three steps: First, we compute the cumulative gain as the sum of the ground truth relevance scores $\hat{r}_\functionloc$ of all retrieved functions $\functionloc\in F$. This ensures the first requirement. To account for the second requirement, we reduce each relevance score proportional to the rank $i$ of function $\functionloc$ using a discount function (e.g., the binary logarithm~\cite{WanWanLiHe13}).
Finally, we normalize the resulting gain to account for stack traces with different lengths. Therefore, we calculate the ideal discounted cumulative gain $\text{IDCG}_k$ \change{as in the previous two steps but assume a perfect ranking (i.e., all relevant functions are positioned at the top ranks).}
\noindent
Formally, the performance measure is thus defined as
$$
\text{NDCG}_k \coloneqq  \frac{1}{\text{IDCG}_k} \sum_{i=1}^{k} \frac{\hat{r}_i}{\text{log}_2(i+1)} \,.
$$
\cstart 
A perfect $NDCG_k$ score of 1 indicates that the relevant functions were returned at the top ranks and a score of 0 implies that no relevant function was among the $k$ retrieved ones. 
Using such a ranking-based measure rather than the scores directly enables a more robust comparison between different scoring methods. In particular, in this case, it does not matter whether one method generally assigns higher scores than another.
\cend

\boldpar{Matching policies}
So far, we were operating under the implicit assumption that all functions in a given stack trace are relevant for a particular crash.
However, this assumption is often inaccurate. In reality, it is generally challenging to pinpoint the exact \emph{root cause} of a crash~\cite{BlaSchAscAbb+20}. Consequently, we often lack precise knowledge about the specific function responsible for introducing a bug. Additionally, crashes might manifest only through particular sequences of function calls, and may not result from a single flawed function~\cite{ZhuBöh21}.

To account for this uncertainty, we opt for a middle ground by over-/ and underestimating our gain. Therefore, we introduce two matching policies:

\italicpar{Optimistic matching} First, we consider an optimistic matching policy that assigns a relevance score of 1 only to the \emph{first} retrieved function from the stack trace. We denote this with $\text{NDCG}^+$. As a consequence, this \emph{overestimates} the performance of a target selection method and serves as a upper bound in our analysis.

\italicpar{Pessimistic matching} Second, we consider a pessimistic policy that assigns a relevance score of 1 to \emph{all} retrieved functions from the stack trace. We denote this with $\text{NDCG}^-$. This underestimates a target selection method and serves as a lower bound.

\subsection{Crash Dataset}
\label{sec:dataset}
\begin{figure}[t]
	\centering
	\includegraphics[width=.85\linewidth]{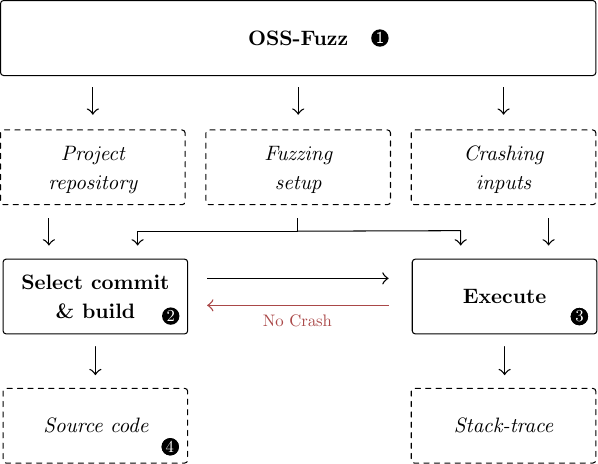}
	\caption{\textbf{Dataset generation process.} \normalfont As basis for our analysis, we collect $1,621$ reproducible crashes. We crawl \ossfuzz, which yields a crashing input and a fuzzing configuration for a project (\stepone). To reproduce the crash, we search for a commit of the project which crashes (\steptwo) when executed under the input from \ossfuzz{} (\stepthree ). Once we reproduce a crash, we extract functions from the project's code and label them according to the stack trace (\stepfour).}
	\label{fig:dataset-process}
\end{figure}

So far, we discussed how we can conceptualize and evaluate target selection methods from an information retrieval perspective. The only missing piece for our evaluation is a ground truth dataset composed of code locations annotated based on whether bugs are present or absent there.

A possible approach to obtain such data points is to fuzz various open-source projects, collect information about identified bugs, and trace back their location within the software project. Unfortunately, this process is very time-consuming and there is no guarantee that we will find all bugs in a given project, which would introduce false negatives to our analysis. As an alternative, we opt to establish our ground truth data by leveraging historical fuzzing campaigns, and, more precisely, crashes detected by the \ossfuzz{}~\cite{Ser+23} project. \ossfuzz{} has consistently applied fuzz testing to an extensive array of open-source projects since 2016 and publicly discloses their findings. While there is still no guarantee that OSS-Fuzz identified every bug, the massive amount of time spent on fuzzing these targets is likely to find a large majority of crashes reachable by a fuzzer.

For using crashing inputs identified from fuzzing for our ground truth, we require two things: (1) Crashes need to be reproducible to verify that they are correct, and (2) we need information about which specific functions are involved in the crash. A good approximation for this is provided by the stack trace of the crash.
Thus, to utilize the data provided by \ossfuzz{}, we must address two primary challenges: First, \ossfuzz{} only releases the time at which a bug is reported and sometimes provides a regression range for when it was resolved but the exact commit of the software project is not disclosed. Second, and more importantly, \ossfuzz{} only shares detailed findings with authorized individuals (i.e., the project maintainers). In particular, the stack trace of a crash is not publicly accessible.

To remediate both issues, we pinpoint a commit at which the crash occurs, which in turn enables us to gather the relevant information for our ground truth. The details of this process are visually depicted in Figure~\ref{fig:dataset-process} and further described in the following.

\begin{figure}
	\begin{center}
		\begin{tikzpicture}
	\begin{axis}[
		ybar,
		ylabel={\#Crashes},
		symbolic x coords={
			\small{Heap overflow},
			\small{ASan seg. fault},
			\small{Memory leak},
			\small{Abort},
			\small{Deadly signal},
			\small{Sig. int. overflow},
			\small{Out of memory},
			\small{Unitialized value},
			\small{Stack overflow},
            \small{UBSan seg. fault},
		},
		xtick=data,
		x tick label style={rotate=45,anchor=east},
		height=5cm,
		width=\linewidth,
		grid=major,
		axis x line*=bottom,
		axis y line*=left,
		nodes near coords,
		nodes near coords align={vertical},
		every node near coord/.append style={font=\tiny},
		legend cell align={left},
	]
		\addplot[coldefault!70, fill=coldefault!70] coordinates {
  			(\small{UBSan seg. fault}, 67)
			(\small{Stack overflow}, 75)
			(\small{Unitialized value}, 88)
			(\small{Out of memory}, 95)
			(\small{Sig. int. overflow}, 103)
			(\small{Deadly signal}, 107)
			(\small{Abort}, 142)
			(\small{Memory leak}, 155)
			(\small{ASan seg. fault}, 174)
			(\small{Heap overflow}, 198)
		};
	\end{axis}
\end{tikzpicture}
	\end{center}
	\caption{\textbf{Crash types.} \normalfont We show the top ten crash types in our ground truth dataset.}
	\label{fig:crash-types}
\end{figure}
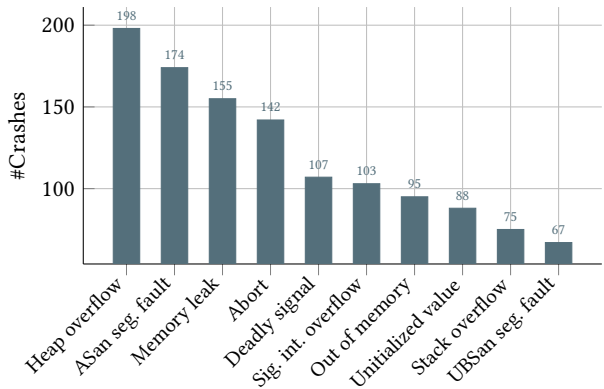

\boldpar{Data collection} 
We start our collection process by scraping the \ossfuzz{} issue tracker, which holds information about each crash, such as the project, reporting time, used fuzzer, sanitizer, and the crashing input~(\stepone). We then select a commit close in time to the initial crash detection~(\steptwo). We rollback the project to this commit and build it with the configuration used by OSS-Fuzz to detect the crash. In case the project build fails, we retry the process by selecting an older commit. In some cases, there are broken references to external dependencies that we fix manually. Once a project is successfully built, we execute the crashing input (\stepthree). If the crash can be observed, we collect a stack trace. If the program does not crash, we go back to \steptwo and retry with an older commit.

After successfully recreating the crash and extracting stack trace information, we proceed to extract every function from the project's source code~(\stepfour). Many projects use pre-compiler directives to enable or disable certain features at compile time, which might include or exclude various code sections. Directly using the source code would thus pose a disadvantage to target selection methods working on the source code level: The source code may actually exhibit defective behavior that, however, will never be labeled as such by \ossfuzz{}, as the defective feature is always disabled via pre-compiler directives. To counteract this, we resort to post-pre-compiler code as the function representation for our dataset. For simplicity, we continue to refer to this code as source code.

\boldpar{Regarding the method's relevance oracle}
The previous section's oracle $\relevanceoracle$ can be approximated by utilizing the information gathered from \ossfuzz{}. Specifically, we can use historic stack trace data and the mapping to the extracted source code to assign functions a score based on whether they were part of a stack trace of a crash found by \ossfuzz{}. This enables us to assemble a labelled corpus that we can further use to compare the retrieval performance of target selection methods to each another.

\begin{figure}
	\begin{center}
		\begin{tikzpicture}
	\begin{axis}[
		xlabel={Traceback length},
		ylabel={\#Tracebacks},
		height=.55\columnwidth,
		width=.95\columnwidth,
		axis x line*=bottom,
		axis y line*=left,
		grid=major,
	]
		\addplot [coldefault!70] table {data/traceback_length.dat};
	\end{axis}
\end{tikzpicture}
	\end{center}
	\caption{\textbf{Stack trace lengths.} \normalfont We show the frequency of stack trace lengths from reproduced crashes. We observe the dominant peak at 7 functions per stack trace.}
	\label{fig:traceback-length}
\end{figure}
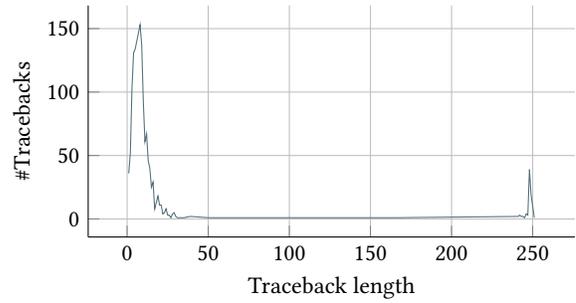

\begin{figure}
	\begin{center}
		\begin{tikzpicture}
	\begin{axis}[
		xbar stacked,
		xmin=0,
		xlabel={\#Crashes},
		hide y axis,
		axis x line*=bottom,
		bar width=1.5,
		height=2.25cm,
		width=\linewidth,
		legend style={
			at={(0.5,2)},
			draw=none,
			anchor=north,
			legend columns={-1},
			column sep=4pt
		},
	]
		\addplot[coldefault!70, pattern=north east lines] coordinates {
		(987,0)
		};
		\addplot[coldefault!70, pattern=crosshatch dots] coordinates {
		(506,0)
		};
		\addplot[coldefault!70, pattern=crosshatch] coordinates {
		(128,0)
		};
		\legend{Address,Memory,Undefined}
	\end{axis}
\end{tikzpicture}
	\end{center}
	\caption{\textbf{Number of crashes per sanitizer.} \normalfont We show the number of crashes in the corpus broken down by sanitizers.}
	\label{fig:sanitizers}
\end{figure}
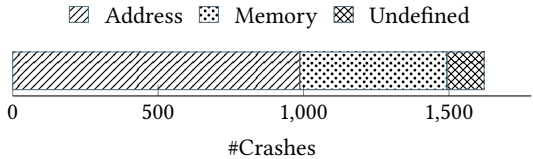

\boldpar{Dataset statistics} We have successfully reproduced 1,621 crashes across 97 C/C++ projects discovered by \ossfuzz{} between 2016 and 2023. We categorize crashes into 48 distinct crash types based on the report by the sanitizer used for identification. Figure \ref{fig:crash-types} provides an overview of of the top ten crash types that contribute for approximately 75\% of the crashes in the corpus, with heap overflows being the most prominent class.
The distribution of crashes in the corpus, broken down by sanitizers, is further illustrated in Figure~\ref{fig:sanitizers} showing that most crashes were identified by address sanitizer, which in total accounted for 62\% of the crashes. The remainder of the crashes are identified by memory sanitizer and undefined behavior sanitizer, which make up 31\% and 7\% of the crashes respectively. Finally, we present the frequency distribution of stack trace lengths in Figure~\ref{fig:traceback-length}. We find the traceback length frequency shows a main peak at seven and a second, less pronounced, peak at 248. While a diverse set of crashes contributes to the former, the latter mostly consists of stack overflow crashes, that is, errors caused by reaching the stack limit.

The dominance of specific crash types and sanitizers used to identify the crashes should be taken into account for the comparison of the target selection methods. 

\section{Comparison of Target Selections}
\label{sec:results}
With our analysis framework at hand, we can now dive into the comparison of target selection methods.
Before we get started, however, we have to make a choice of the methods which we take into account.

\subsection{Choice of Target Selection Methods}
\label{sec:target-selection-methods}
Before a fuzzer can be directed towards a specific location, we must make a suitable choice in terms of target selection. With Table~\ref{tab:survey} outlining the common techniques used for fuzzing, we have ample choices to select from. In the following, we desire to evaluate the quality of target selection methods. For an informed evaluation, we select three target selection approaches covering both core methods, pattern-based and metrics-based methods: For the former, we pick up the idea of using sanitizer instrumentation \cite{OstRazBosGiu+20, CheLiXuGuo+20, ZheZhaHuaRen+23} and the idea of using recently modified code changes \cite{ZhuBöh21} for target selection. For the latter, we resort to the code metrics presented by Leopard~\cite{DuCheLiGuo+19}.

To better put these methods used by fuzzing into context, we also include approaches no fuzzing paper has used yet.
First, we turn to \emph{static analysis}, which has found widespread adoption in industry~\cite{FacebookInfer, GoogleSAST, GithubSAST}.
% ~\todo{cite Facebook's Infer and some other tools widely used. Lukas marked as DONE (?): cited tech reports of SAST usage in the industry (FB, Google, Github)}
Essentially, any static software application security testing (SAST) tool can be converted into a target selection method, with the potentially faulty lines as target locations. A fuzzer can then be used to create a proof of concept, which static analysis usually cannot do on its own. 
For our analysis, we choose two SAST tools, Rats~\cite{rats} and Cppcheck~\cite{cppcheck}. 
Initially developed by Secure Software Inc. in 2001, Rats is one of the earliest freely available SAST tools and although its development was discontinued in 2013, it stayed relevant as a  common academic baseline~\cite{SASTeval2018, KAUR20202023, 10.1145/3475716.3475781}.
Cppcheck is a mature open-source SAST tool with a very active community continuously developing and extending the software. 
Its availability in package repositories of many major Linux distributions, precise error messages, integration with many popular IDEs and its ease of use makes it a popular choice for initial security assessments of C/C++ code~\cite{cppcheck, 8531281}.
% \todo{there must be some explanation why these two in particular (something like: poplular / widely used)}
Second, we want to capture the broader trends in other areas of code assessment, which rapidly turn to learning-based approaches. In particular, we select three source code based vulnerability prediction models for this purpose: ReVeal~\cite{ChaKriDinRay+21}, which is based on graph neural networks, Linevul~\cite{FuTan+22}, resorting to a encoder-only transformer based architecture, and a fine-tuned version of the transformer based language model CodeT5+~\cite{wang2023codet5plus}.

Finally, we include one target selection that uses a random scoring of code locations as a baseline for the other models.
In the following, we provide a brief overview of our selected methods.

\boldpar{CodeT5+}
Based on the transformer architecture, CodeT5+~\cite{wang2023codet5plus} is a large language model for programming tasks. It is pre-trained on a subset of the CodeSearchNet~\cite{codesearchnet} and GitHub Code datasets\footnote{https://huggingface.co/datasets/codeparrot/github-code}.
Pre-training consists of a first stage with unimodal tasks such as denoising, followed by a second stage using text-code bimodal data,  aiming to improve code generation and understanding tasks. For our experiments, we use the 220 million parameter version and extend it to perform  a vulnerability prediction task similarly to Chen et al.~\cite{CheDinAloChe+23}. To this end, we complement the model with a prediction head for binary classification as described in~\cite{brokenPromises} and fine-tune it on the DiverseVul dataset~\cite{CheDinAloChe+23}, consisting of ca. 350k (19k vulnerable) annotated methods representing 150 CWEs.

% SAST
\boldpar{Cppcheck}
The Cppcheck SAST tool is able to perform a variety of checks on C and C++ source code, scanning for undefined behavior and dangerous code patterns that might indicate vulnerabilities~\cite{cppcheck}. 
Checks include analysis passes for automatic variable checking, array bounds checking or dead code elimination.  
A recent study evaluating the tool on Mozilla Firefox found that Cppcheck was able to find 83.5\%  of the vulnerabilities with only 7.2\% false positives~\cite{cppcheckStudy}.
\change{Cppcheck can be used with different pre-built configurations. We use the default setup which includes all of them.}

\boldpar{Rats}
The \textit{Rough Auditing Tool for Security} (Rats)~\cite{rats} is one of the earlier rule-based SAST solutions. The tool offers language support for C/C++, Perl, PHP, Python and Ruby with varying degrees of maturity; while it is able to scan C code for more complex bug classes such as Time-of-Check-Time-of-Use,  analysis capabilities in Python are constrained to checking for potentially dangerous built-in or library function calls. 
\change{The results of Rats can be filtered by three types of severity, which we include all. Besides that, we use the default configuration.}

\boldpar{Leopard-C}
Du et al. propose Leopard as a framework for identifying potentially vulnerable code sites in C/C++ programs~\cite{DuCheLiGuo+19}.
The framework primarily works in two stages, the first one employing code complexity metrics, hereinafter referred to as \textit{Leopard-C}, to sort functions into different bins, and the second one using vulnerability metrics (\textit{Leopard-V}) to rank functions inside those bins. 
Leopard-C takes into account the cyclomatic complexity of a function, i.e., the number of linearly independent Control Flow paths, and a set of loop metrics: the number of loops and nested loops as well as the maximum nesting level.
The intuition is that more complex functions are harder to analyze and reason about.

\boldpar{Leopard-V}
With Leopard-V, the authors derive a new set of metrics from common vulnerability causes~\cite{DuCheLiGuo+19}. 
\change{More precisely, they select eleven code metrics that aim to capture the following characteristics: the dependency to other functions, the use of pointers, and features of employed control structures. For the first characteristic, they count the number of parameters of the function and the number of parameters to its callees. To measure the use of pointers, they, for example, count the number of variables involved in pointer arithmetic. Lastly, they measure the features of the control structures such as the number of nesting levels or the number of conditional statements without an alternative. Finally, Du et al. sum up all individual metrics into a single vulnerability metric.}

% Heuristics
\boldpar{Sanitizer instrumentation}
Our sanitizer based selection methods builds on the idea of ParmeSan~\cite{OstRazBosGiu+20}, SAVIOR~\cite{CheLiXuGuo+20}, and FishFuzz~\cite{ZheZhaHuaRen+23}, which utilize information about the code regions augmented with instrumentation by the employed sanitizers. This is based on the rationale that sanitizer instrumentation is added at code locations where defective behavior might occur (e.g., array accesses).
\change{To implement this, we count the number of sanitizer callbacks added to each function. These callbacks are added during compilation, for example, when memory is allocated or freed, or other memory access that may potentially allow for errors are detected. Consequently, a function with many sanitizer callbacks indicates a higher likelihood for a sanitizer to detect an error in this function than in functions with very few or none added callbacks.}

\boldpar{ReVeal}
ReVeal~\cite{ChaKriDinRay+21} is a Graph Neural Network (GNN) for vulnerability detection based on Zhou et al.'s approach of learning program semantics~\cite{ZhoLiuSioDu+19}.
It uses Code Property Graphs, a holistic code representation combining Abstract Syntax Trees with Dataflow- and Control Flow Graphs~\cite{YamGolArpRie+14}, as an input to the eight-step GNN and augment it with a classification layer to separate the learning of code representations from the learning of vulnerability indicators.
The model is trained on a large real-world dataset of vulnerabilities in the Chromium and Debian Kernel sources. 
We use the implementation from~\cite{revealGH}.

\boldpar{Linevul}
Using Microsoft's CodeBERT~\cite{feng-etal-2020-codebert}, a 125M parameter encoder-only transformer for programming tasks, as a foundation model, Linevul~\cite{FuTan+22} provides line-level vulnerability prediction for C/C++ code. 
Following results from~\cite{brokenPromises}, we use the original model trained on the BigVul dataset~\cite{bigvul} but apply normalization to any code input before inference in order to reduce confounding introduced by code style variations. 

\cstart
\boldpar{Recently modified code (``Recent'')}
Several directed fuzzing approaches resort to recently modified code locations as their targets~\cite{BoePhaNguRoy17, ZhuBöh21, CanMatGraKal+22, PenLiLiuXu+19}. \citet{ZhuBöh21} even find that most (four out of five) bugs found by \ossfuzz{} are introduced by recent code changes. Therefore, we also include such a target selection in our experiments. To this end, we first assign each function the time it was last modified. We identify the changes based on information from the version tracking system employed by the individual projects. When every function has been assigned a timestamp, we rank them with the most recently changed functions ranking highest.
\cend

\boldpar{Random}
To provide a baseline for the previous methods, we also include a target selection which outputs a random ranking over all code locations.

\subsection{Results and Insights}
\begin{figure}[t]
	\centering
	\includegraphics[width=0.9\linewidth]{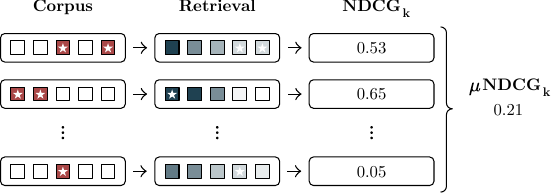}
	\caption{\textbf{Retrieval and evaluation process.} \normalfont Our crash corpus consists of various projects with functions labelled as relevant (\raisebox{-0.15em}{\FiveStarOpen}) or irrelevant. A target selection method is used to create a ranking of the functions with the $k$ highest ranks forming the retrieval. For each retrieval we calculate the NDCG and average the results across all crashes in every project of our corpus, leading to the $\mu\text{NDCG}_k$ as the retrieval score for a selection method on $k$ sized retrievals.} 
	\label{fig:retrieval-process}
\end{figure}

\cstart
\begin{figure*}[t]
	\begin{center}
		\begin{tikzpicture}
\begin{groupplot}[
		group style={
			group name=ndcgplot,
			group size= 2 by 1,
			horizontal sep=2cm,
			vertical sep=1.70cm
		},
		height=4.5cm,
		width=8cm,
		grid=major,
		yticklabel style={/pgf/number format/fixed},
		legend cell align={left},
		ymin=-0.02,
		ymax=0.32,
	]
	\nextgroupplot[
		ylabel={$\mu\text{NDCG}^-$},
		xlabel={\#Retrieved functions},
		legend columns=5,
		legend style={
			at={(.7\textwidth, -0.4)},
			column sep=4pt,
		},
		title={a) Under-estimating},
		xmode=log,
		axis x line*=bottom,
		axis y line*=left,
	]
		\addplot [const plot, colcodet5p] table {data/mean-ndcg-ue-codet5p-1000-nofilter.dat};
		\addlegendentry{CodeT5+};
		\addplot [const plot, dashed, colreveal] table {data/mean-ndcg-ue-reveal-1000-nofilter.dat};
		\addlegendentry{ReVeal};
		\addplot [const plot, dashdotted, collinevul] table {data/mean-ndcg-ue-linevul-1000-nofilter.dat};
		\addlegendentry{Linevul};
		\addplot [const plot, colrats] table {data/mean-ndcg-ue-rats-1000-nofilter.dat};
		\addlegendentry{Rats};
		\addplot [const plot, dashed, colcppcheck] table {data/mean-ndcg-ue-cppcheck-1000-nofilter.dat};
		\addlegendentry{Cppcheck};
		\addplot [const plot, dashed, colcomplexity] table {data/mean-ndcg-ue-complexity-1000-nofilter.dat};
		\addlegendentry{Leopard-C};
		\addplot [const plot, colvulnerability] table {data/mean-ndcg-ue-vulnerability-1000-nofilter.dat};
		\addlegendentry{Leopard-V};
		\addplot [const plot, colsan] table {data/mean-ndcg-ue-sanitizer-1000-nofilter.dat};
		\addlegendentry{Sanitizer};
        \addplot [const plot, colrecent] table {data/mean-ndcg-ue-recent-1000-nofilter.dat};
		\addlegendentry{Recent};
		\addplot [const plot, colrandom] table {data/mean-ndcg-ue-random-1000-nofilter.dat};
		\addlegendentry{Random};
	\nextgroupplot[
		ylabel={$\mu\text{NDCG}^+$},
		xlabel={\#Retrieved functions},
		title={b) Over-estimating},
		axis x line*=bottom,
		axis y line*=left,
		xmode=log,
	]
		\addplot [const plot, colcodet5p] table {data/mean-ndcg-oe-codet5p-1000-nofilter.dat};
		\addplot [const plot, dashed, colreveal] table {data/mean-ndcg-oe-reveal-1000-nofilter.dat};
		\addplot [const plot, dashdotted, collinevul] table {data/mean-ndcg-oe-linevul-1000-nofilter.dat};
		\addplot [const plot, colrats] table {data/mean-ndcg-oe-rats-1000-nofilter.dat};
		\addplot [const plot, dashed, colcppcheck] table {data/mean-ndcg-oe-cppcheck-1000-nofilter.dat};
		\addplot [const plot, dashed, colcomplexity] table {data/mean-ndcg-oe-complexity-1000-nofilter.dat};
		\addplot [const plot, colvulnerability] table {data/mean-ndcg-oe-vulnerability-1000-nofilter.dat};
		\addplot [const plot, colasan] table {data/mean-ndcg-oe-sanitizer-1000-nofilter.dat};
        \addplot [const plot, colrecent] table {data/mean-ndcg-oe-recent-1000-nofilter.dat};
		\addplot [const plot, colrandom] table {data/mean-ndcg-oe-random-1000-nofilter.dat};
\end{groupplot}
\end{tikzpicture}
	\end{center}
	\caption{\textbf{Mean retrieval scores of the target selection methods.} \normalfont Mean NDCG scores of the target selection methods across the whole corpus for the first $k$ retrieved functions. Target selection methods belonging to the same general class of approaches, i.e., SAST tools, vulnerability prediction models, code metrics, and sanitizer instrumentation-based share the same color.}
	\label{fig:mean-gains}
\end{figure*}
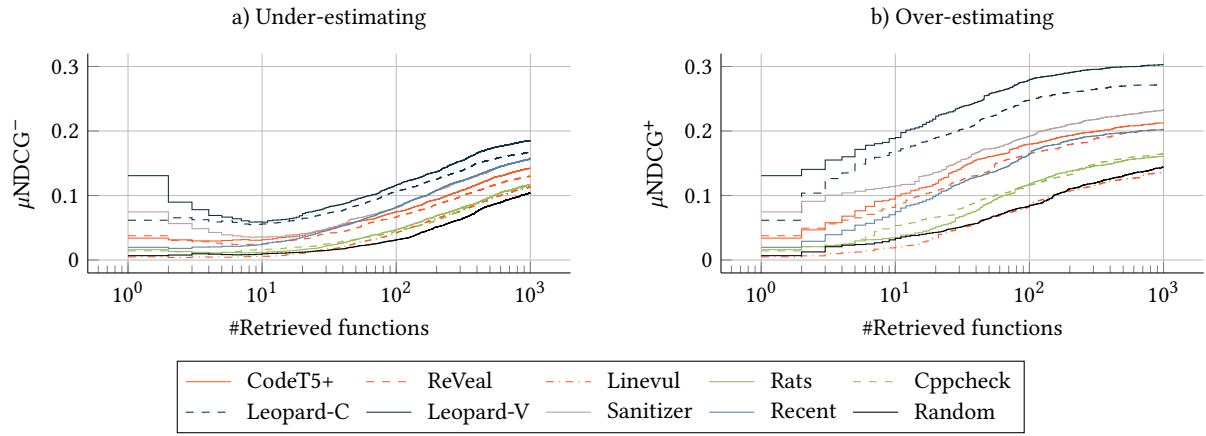
\cend

As discussed in Section \ref{sec:methodology}, the problem of finding suitable targets in a project to guide a fuzzer towards can be framed as an information retrieval problem. As such, we approach the evaluation of the target selection methods with a standard metric to measure the quality of retrieval algorithms with, the NDCG.

In Section \ref{sec:dataset}, we described the assembly of our dataset based on information from \ossfuzz{}.
By reproducing crashes of projects in defective versions, we gathered stack traces which we used
to assign a relevance score to each function in the respective project. This way we ended up with a corpus
of projects in specific versions and functions labelled as relevant or irrelevant, based on whether they appeared in the stack trace of the crash, or not. This is shown in Figure \ref{fig:retrieval-process}.  We understand target selection as the retrieval of these relevant functions for each project in a defective version. To that end, the functions are ranked by using a target selection method with the $k$ highest ranking function forming the retrieval. We calculate the NDCG for each of these retrievals and take the mean of the NDCG score, $\mu\text{NDCG}_k$, to quantify the retrieval performance of the target selection method.

As our labeling is based on crash stack traces and not every function in a given stack trace is defective, we have to deal with a certain degree of label noise, which we counteract by using one metric that under-estimates and one
that over-estimates the true quality of our selection methods, the $\text{NDCG}^-$ and the $\text{NDCG}^+$, respectively, as described in Section \ref{sec:methodology}.

In this section, we compare the retrieval performance of the target selection methods. As noted in Section~\ref{sec:dataset}, different crash types do occur with different frequencies and are, hence, included in our dataset with varying amounts. The same holds for crashes found by the three sanitizers that we took into account for the crash reproduction. This makes it necessary to take a closer look at the target selection methods with regard to their quality for those categories in our dataset.
However, before examining these specific cases, we begin by taking the general retrieval performance into the focus.

%%%
%%% General performance
%%%
\boldpar{General performance}
The performance of the target selection methods is shown in Figure
\ref{fig:mean-gains}. To get an intuition of how to interpret the $\mu\text{NDCG}_k$ scores, recall that the NDCG expresses the degree to which the ranking
created by a selection method matches an ideal ranking. An ideal ranking would exclusively assign functions from a crash stack trace to the first ranks; thus, an ideal NDCG score of 1.0 would
express that a given target selection algorithm is able to match this ranking and
thus predict crashing functions perfectly.

One particularity we observe in the case of the under-estimating $\text{NDCG}^-$ is the drop in retrieval performance for for $k > 1$ retrieved functions for Leopard-V, as can be seen in Figure \ref{fig:mean-gains} a).  When considering solely the top-ranking function (i.e.,\; $k=1$), the Leopard-V method exhibits a NDCG score of 0.13. This implies that the initial function in the ranking of this method aligns with that of an ideal algorithm in 13\% of instances. Put differently,
the highest ranking function of Leopard-V is part of a crash's stack trace in 13\%
of the cases across every crash in our corpus. With an increasing number of retrieved functions, i.e., greater values for $k$ in the plot, the under-estimating $\text{NDCG}^-$ drops, which
shows that the target selection method does not continue to rank the other
functions from the stack trace as highly. When taking into account that the
under-estimating matching policy assigns an equal relevance to every function from
the stack trace, this indicates, that only one of the stack trace's functions is
considered very likely to be defective by the selection method. This could be explained
by the fact that in many cases only one function from the stack trace actually is
defective in practice.
This behavior does not arise in the over-estimating case, as in this
case only the highest ranking function from the stack trace is assigned a relevance~$\relevancescore > 0$.

% Statistics brief description
In order to further investigate the relationship of the respective methods to each other, we conducted a Mann-Whitney-U test to identify significance\footnote{We consider $p < 0.05$ statistical significant. We publish p-values together with our source code.}. Furthermore, we refer to a method as \textit{dominant} over another, if its distribution of $\text{NDCG}_k$ scores is significantly higher for every number $k$ of retrieved functions.

\cstart
Let us briefly summarize our insights in this regard. First, we shift our focus to the case in which at least 25 functions were retrieved ($k \geq 25$) as this marks the point from which on every method, except for Linevul, significantly outperforms the random baseline. In this case, we find four clusters of target selection methods which can be recognized particularly well in the over-estimating plot in Figure \ref{fig:mean-gains} b). The clusters are characterized by the fact that the methods within a cluster do not dominate one another, but each method of one cluster dominates each method of the one beneath. The first cluster is formed by Leopard-V and Leopard-C which outperform every other method. The second one consists of the sanitizer-based method, CodeT5+, ReVeal, and the recent code changes heuristic. This is followed by the SAST tools cluster containing Rats and Cppcheck which still dominate the last cluster, Linevul and the random scoring. The original Linevul model likely performs similarly to the random baseline as it picks up spurious correlations with code formatting~\cite{brokenPromises}.

When we shift our view to less than 25 retrieved functions ($k < 25$), we find a different picture. We observe that the sanitizer-based method performs better than the other three methods with which it forms a cluster for $k \geq 25$ for a few number of returned functions. More precisely, it significantly outperforms the next best methods CodeT5+ and ReVeal for $k < 5$. Also, for the same interval, the method based on how recently a code location was last modified performs significantly worse than the next best methods CodeT5+ and ReVeal. This shows how methods that do not significantly differ for a larger number of retrieved functions can perform vastly different when queried for a few number of relevant functions.

To understand the practical implications of these results, we take a closer look at the case of AFLChurn \cite{ZhuBöh21}. In their case study the authors compare their proposed directed fuzzer, AFLChurn, which can operate on targets from a continuous target selection method, to AFLGo, which requires a discrete one. For the target selection they assign each code locations a continuous score based on how recently it was modified. Consequently, AFLGo, requiring a discrete set of targets, must decide on a threshold score or a maximum amount of code locations to select as targets for the fuzzing process.
When measuring the time to expose a crash in libhtb, one of their target programs, the authors find that AFLGo takes 3 hours and 12 minutes on average while AFLChurn clocks in at 2 hours and 1 minute. Now, our experiments add a new angle to explain this performance gain. We observe that a target selection based on how recently a code locations was last modified only picks up performance with an increasing number of retrieved functions compared to the sanitizer or even code metric-based methods. Consequently, in this case a fuzzer that requires a discrete selection of targets and must therefore decide on a threshold score or maximum amount of targets cannot profit as well as one operating with a continuous target selection method can.

We hypothesize that the difference of the approaches would have been less pronounced for a method that also performs well for a low number of retrieved functions, such as Leopard-V, a sanitizer-based method or Leopard-C. Also, we conjecture that a target selection approach should always be chosen in accordance with the requirements of the fuzzing approach. For example, machine learning-based approaches, such as our fine-tuned CodeT5+ or ReVeal, might work well for fuzzers that can operate on continuous target selections, while showing sub-optimal performance for those that cannot. However, more research is necessary in this direction to better substantiate these hypotheses. % with empirical evidence.

\cend

\begin{tcolorbox}[boxrule=1pt, arc=1mm, colback=white, colframe=coldefault]
	Target selection methods based on software metrics perform significantly
	better than every other considered method. The best software metric,
	Leopard-V, correctly captures as much as 13\% of the crashes with its
	highest ranking function across the whole corpus of more than 1600 crashes. This makes it the most natural and really only viable candidate for
	  fuzzing approaches which require a discrete selection method.
\end{tcolorbox}

\boldpar{Impact of sanitizers}
Our corpus consists of twice as many bugs found by address sanitizer than memory sanitizer or undefined behavior sanitizer combined. Therefore, only taking the retrieval performance on the whole corpus into account largely reflects the performance to retrieve functions from stack traces of crashes found by address sanitizer. This potentially conceals good performance measures with regard to crashes of one of the less frequent sanitizers.
\cstart
The individual performances of the target selection methods broken down by sanitizers are shown in Appendix~\ref{sec:appx-break-down-sanitizers}. They further support the dominance of Leopard's methods across most numbers of retrieved functions. However, especially for the crashes discovered by the memory sanitizer, the sanitizer-based target selection method works almost as well.
\cend

\begin{figure}
	\begin{center}
		\begin{tikzpicture}
\begin{groupplot}[
		group style={
			group name=ndcgplot,
			group size= 1 by 2,
			horizontal sep=2cm,
			vertical sep=1.75cm
		},
		height=4.2cm,
		width=8cm,
		grid=major,
		yticklabel style={/pgf/number format/fixed},
		legend cell align={left},
	]
   	\nextgroupplot[
		title={a) Leopard-V broken down by sanitizers},
		xlabel={\#Retrieved functions},
		ylabel style={align=center},
		ylabel={$\mu\text{NDCG}^-$},
		grid=major,
		axis x line*=bottom,
		axis y line*=left,
		legend pos=north west,
		legend cell align={left},
		xmode=log,
        ymin=-0.02,
        ymax=0.32,
	]
		\addplot [const plot, dashed, colvulnerability] table {data/mean-ndcg-ue-vulnerability-1000-nofilter.dat};
        
        % Best
        \addplot [const plot, colvulnerability!51] table {data/mean-ndcg-ue-vulnerability-1000-sanitizer=memory.dat};
        
        \addplot [const plot, colvulnerability!] table {data/mean-ndcg-ue-vulnerability-1000-sanitizer=address.dat};
        \addplot [const plot, colvulnerability!13] table {data/mean-ndcg-ue-vulnerability-1000-sanitizer=undefined.dat};
	\nextgroupplot[
		title={b) Leopard-V broken down by crash types},
		xlabel={\#Retrieved functions},
		ylabel style={align=center},
		ylabel={$\mu\text{NDCG}^-$},
		grid=major,
		axis x line*=bottom,
		axis y line*=left,
		legend pos=north west,
		legend cell align={left},
		xmode=log,
        ymin=-0.02,
        ymax=0.32,
	]
		\addplot [const plot, dashed, colvulnerability] table {data/mean-ndcg-ue-vulnerability-1000-nofilter.dat};
		% 198
		\addplot [const plot, colvulnerability!100] table {data/mean-ndcg-ue-vulnerability-1000-crash_class=asan_heap_buffer_overflow.dat};
		% 174
		\addplot [const plot, colvulnerability!88] table {data/mean-ndcg-ue-vulnerability-1000-crash_class=asan_segv.dat};
		% 155
		\addplot [const plot, colvulnerability!78] table {data/mean-ndcg-ue-vulnerability-1000-crash_class=leaksan_detected_memory_leaks.dat};
		% 142
		\addplot [const plot, colvulnerability!71] table {data/mean-ndcg-ue-vulnerability-1000-crash_class=asan_abrt_unknown_address.dat};
		% 107
		\addplot [const plot, colvulnerability!54] table {data/mean-ndcg-ue-vulnerability-1000-crash_class=libfuzzer_deadly_signal.dat};
		% 103
		\addplot [const plot, colvulnerability!52] table {data/mean-ndcg-ue-vulnerability-1000-crash_class=ubsan_runtime_error_signed_integer_overflow.dat};
		% 95
		\addplot [const plot, colvulnerability!47] table {data/mean-ndcg-ue-vulnerability-1000-crash_class=libfuzzer_out_of_memory.dat};
		% 88
		\addplot [const plot, colvulnerability!44] table {data/mean-ndcg-ue-vulnerability-1000-crash_class=msan_use_of_uninitialized_value.dat};
		% 75
		\addplot [const plot, colvulnerability!38] table {data/mean-ndcg-ue-vulnerability-1000-crash_class=asan_stack_overflow.dat};
		% 67
		\addplot [const plot, colvulnerability!34] table {data/mean-ndcg-ue-vulnerability-1000-crash_class=ubsan_segv_on_unknown_address.dat};
		% 52
		\addplot [const plot, colvulnerability!26] table {data/mean-ndcg-ue-vulnerability-1000-crash_class=libfuzzer_timeout.dat};
\end{groupplot}
\end{tikzpicture}
	\end{center}
	\caption{\textbf{Retrieval performance of Leopard-V across different sanitizer and crash types.} \normalfont Each solid line represents a crash type or sanitizer, respectively. The color strength reflects the prevalence the respective sanitizer and crash type in the dataset. The performance across the whole dataset is added in dashed, for reference.}
	\label{fig:mean-gains-leopard}
\end{figure}
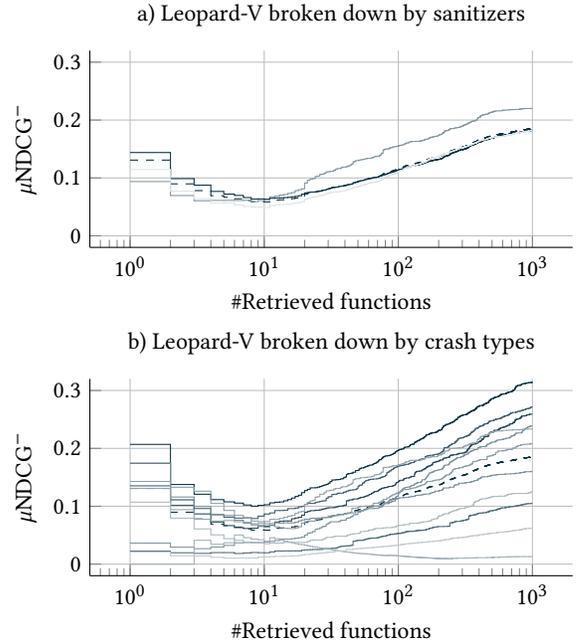

To further inspect the variance in retrieval performance for an individual selection method, we exemplary show 
the retrieval quality of Leopard-V for the three sanitizers in Figure \ref{fig:mean-gains-leopard}a), indicating
that the impact of which sanitizer was used to find a crash on the retrieval performance is limited.  

%%%
%%% Performance by crash type
%%%
\boldpar{Impact of crash types}
Just like with the sanitizers, we can conduct a finer grained assessment of the selection methods performance by breaking them down with respect to the crash types. \change{The resulting retrieval performances are depicted in Appendix \ref{sec:appx-break-down-crashtypes}. We observe that the Leopard methods, again, perform best across all examined crash types with the gap to the next best method, the sanitizer-based method, being most pronounced for heap buffer overflows and less pronounced for segmentation faults.}

To inspect how strong the retrieval performance depends on the crash type, we take a look at the quality of the method performing strong across all crash types (i.e., Leopard-V) in greater detail. We depict the retrieval performance for every crash type of which we have at least 50 samples in the dataset in Figure \ref{fig:mean-gains-leopard}b). The color strength indicates how prevalent the respective crash type is in our corpus. It shows that even the best performing selection method's retrieval quality varies greatly across the various crash types. \change{Hence, unlike with the sanitizers used to discover the crashes which only have limited impact on the retrieval performance, the quality greatly depends on the crash type. Some crash types, such as heap buffer overflows can be detected very well using the Leopard-V method while others, such as stack buffer overflows, can hardly be detected.}

\cstart
\begin{tcolorbox}[boxrule=1pt, arc=1mm, colback=white, colframe=coldefault]
	Software metric-based target selection methods perform significantly better than any other method across most types of crashes and sanitizers. The only other method close to their performance in some cases is the sanitizer-based one.
\end{tcolorbox}
\cend

\change{
\section{Limitations}
\label{sec:limitations}
In our experiments, we investigate the effectiveness of target selection methods in identifying worthwhile targets for directed fuzzers. For this, we treat target selection as an information retrieval problem and compile a dataset from the source code and tracebacks of various crashes we reproduced from the \ossfuzz{} project.

\boldpar{False positives}
Underlying our analysis is the assumption that a target which is part of a crash's traceback would also have constituted a valuable target for a directed fuzzer. However, there is a semantic gap between a target's occurrence in a traceback and its value for a directed fuzzer, i.e., how well it aids in triggering the root cause of a crash. It is possible that in some scenarios the root cause of the crash is not part of the traceback, for example, in case of intricate data-dependencies which have to be met or for multi-threaded or multi-process applications. As we resort to the traceback for our label source, this could lead to false positive labels in our dataset. Yet, previous work has shown that using the traceback of crashes as targets for a directed fuzzer does in fact reduce the time to expose the crashes compared to an undirected fuzzing approach \cite{CheXueLiChe+18, DuLiLiuMao+22, BoePhaNguRoy17}. This indicates that the traceback is indeed a good approximation of the crash's root cause for this application.

\boldpar{False negatives}
As the ground truth for our evaluation, we resort to crashes found by the \ossfuzz{} project. This poses the risk of missing crashes that have not yet been identified by the project. In particular, individual targets could be erroneously assigned as not being involved in any crash, when in fact the respective crash has just not yet been found by the fuzzers (i.e., a false negative). While this could potentially lead to an underestimation of the actual scoring quality of the target selection methods, the fact that \ossfuzz{} continuously fuzzes each project with >200 cores per project on average\footnote{as of March 2021} alleviates this risk.

\boldpar{Target granularity}
In our experiments, we focus on function level target granularity. While some selection methods and our ground-truth information from the tracebacks would have allowed for a more fine-grained evaluation, for example on line level, this would have excluded several of the other methods. We thus adopted a meet-in-the-middle approach on function level which allows for all target selection methods to be compared to one other while still offering a fine enough granularity to be useful for directed fuzzing. This is shown by the fact that most publications from our literature analysis in Section \ref{sec:localization}, which introduce a new target selection method, use a function level granularity \cite{CaoHeSunOuy+23, ZhaLiaXiaZha+22, ZheZhaHuaRen+23, DuCheLiGuo+19, KruGriRos22}.
}

\change{
\section{Related Work}
\label{sec:related}
Our systematization of target selection methods for directed fuzzers is related to two types of prior works: surveys of fuzzing literature and analyses aimed at enhancing specific components of the fuzzing pipeline.

\boldpar{Literature surveys}
Several surveys on fuzzing research have been conducted throughout the years. Although they are methodologically similar to each other and our work, their respective focus varies. First, there are surveys that take the whole fuzzing pipeline into account. \citet{ManHanHanCha+21} and \citet{LiaPeiJiaShe+18}, for example, both introduce general multi-step models of the fuzzing process and survey existing literature with regard to each step in their model. More closely related to our work is the survey by \citet{WanZhoYueLin+24} who focus on directed fuzzers. To that end, they identify several characteristics of directed fuzzers for which they examine prior works. While most of these characteristics are concerned with how the fuzzer operates, one characteristic covers the method by which its targets are selected. However, as their focus is on the fuzzer itself rather than on its preceding target selection method, they merely identify which method was used but do not examine it any further.

Other fuzzing surveys take a more specific perspective and focus on how certain methods are applied in the fuzzing pipeline or challenges that arise when applying fuzzing to particular application areas. That is, for example, how machine learning techniques are used for fuzzing~\cite{SavRodDun+19, WanJiaLiuHua+20} or the application of fuzzing to find flaws in embedded devices~\cite{MueStiKarFra+18, EisMauShrHut+22}, respectively. 

Lastly, surveys such as those by \citet{SchBarSchBer+24}, \citet{KleRueCooWei+18} or \citet{KimChoImHeo+24} take a meta perspective and study fuzzing research itself. To that end, they examine the process conducted to evaluate fuzzers in various publications. Based on their findings they can derive information about the general validity of the research field as well as recommendations on how to conduct an evaluation ideally.

\boldpar{Enhancing fuzzer components}
In addition to surveying publications on directed fuzzing, we also focus on systematically investigating the step preceding directed fuzzers; namely, the methods employed to select their targets. This is related to prior works which have conducted experiments on individual steps of the fuzzing pipeline. \citet{BöhPhaRoy16}, for example, study various power- and search-strategies to improve the seed scheduling part of a fuzzer, \citet{WuJiaXiaHua+22} compare different setups for a mutation strategy, and \citet{HerGunMagSha+21} focus on the seed selection and compare several different methods for that purpose. In contrast, our work focuses on the target selection, which has not yet been studied in-depth, and is, thus, orthogonal to other improvements.
}

\section{Conclusion}
\label{sec:conclusion}
Our analysis adds a new piece to the puzzle of understanding fuzzing performance.
Previous research in directed fuzzing focused on faster and deeper exploration of targets.
While this challenge continues to hold enough problems for future work, we demonstrate that the targets within the programs also play a crucial role and the effectiveness of directed fuzzing stands or falls with the quality of the targeted code locations. Our work thus helps to decouple orthogonal factors of performance and provides a better understanding of the \emph{where to fuzz} in the discovery of defects.

Furthermore, the results of our analysis provide a surprising insight: Despite extensive research on locating interesting code for fuzzing, classic software metrics still outperform more sophisticated counterparts in target selection. Given the capabilities of some pattern-based methods, this is unexpected and suggests significant room for improvement. In particular, we consider recent machine learning models a promising alternative, as these models can acquire remarkable skills and thus potentially push the performance of target selection forward,  generating a better basis for directed fuzzing in practice.

We make our source code and data publicly available online at \url{https://github.com/wsbrg/crashminer}.

\section{Acknowledgements}
\label{sec:acknowledgements}
This work was funded by Deutsche Forschungsgemeinschaft (DFG, German Research Foundation) under Germany’s Excellence Strategy - EXC 2092 CASA - (390781972), the European Research Council (ERC) under the consolidator grant MALFOY (101043410), and the German Federal Ministry of Education and Research under the grant BIFOLD24B.

% Bibliography
\ifacm
\balance
\bibliography{
    bib/dblp-bib/sp,
    bib/dblp-bib/uss,
    bib/dblp-bib/ndss,
    bib/dblp-bib/icse,
    bib/dblp-bib/ccs,
    bib/dblp-bib/nips,
    bib/additional,
    bib/mlsec-bib/sec
}
\fi
\ifieee
\bibliography{
	IEEEabrv,
	bib/dblp-bib/ccs,
	bib/dblp-bib/raid,
	bib/dblp-bib/esorics,
	bib/dblp-bib/sp,
	bib/dblp-bib/ndss,
	bib/dblp-bib/uss,
	bib/dblp-bib/codaspy,
	bib/additional,
	bib/dblp-bib/eurosp,
	bib/dblp-bib/nips,
	bib/dblp-bib/icse,
    bib/mlsec-bib/sec
}
\fi

% Appendix
\clearpage
\appendix

\section{Additional Experiments}
\cstart
Alongside the impact of the sanitizer and crash type on the performance of the target selection methods, we also analyzed how the traceback lengths of the crashes might influence the results.

While measuring the traceback lengths' impact is seemingly straightforward, it comes with it own hurdles. In particular, some projects, such as libraw and libpcap, differ greatly in their crashes' average traceback length of 5.2 and 12.4, respectively. The same holds for crash types such as the use of an uninitialized value and a stack overflow whose traceback lengths differs by 230 functions on average. This has to be taken into account when evaluating the impact of the traceback length's. Otherwise the performance on individual projects or crash types would indirectly be measured in this assessment.

To account for this, we focus on the mean percent increase/decrease between the longest and shortest traceback for every combination of projects and crash types. We find that there is only a negligible change of the results for the under-estimating NDCG (less than 0.1\%). For the over-estimating NDCG, the impact is higher, but still only shows a limited impact measuring a 2\% performance increase for longer tracebacks. The different behaviour can be explained by the optimistic and pessimistic matching strategies applied for the over- and under-estimating NDCG. While the former rewards if \emph{any} function from the traceback was ranked highly, the latter one expects \emph{every} function from the traceback to be at high positions. Highly ranking any function is significantly simpler for longer tracebacks which could explain the impact of the traceback length.

\cend

% \onecolumn
\section{Retrieval Performance by Sanitizers}
\label{sec:appx-break-down-sanitizers}

\vspace{2em}

% \begin{figure*}
\begin{center}
    \includegraphics[width=1.00\linewidth]{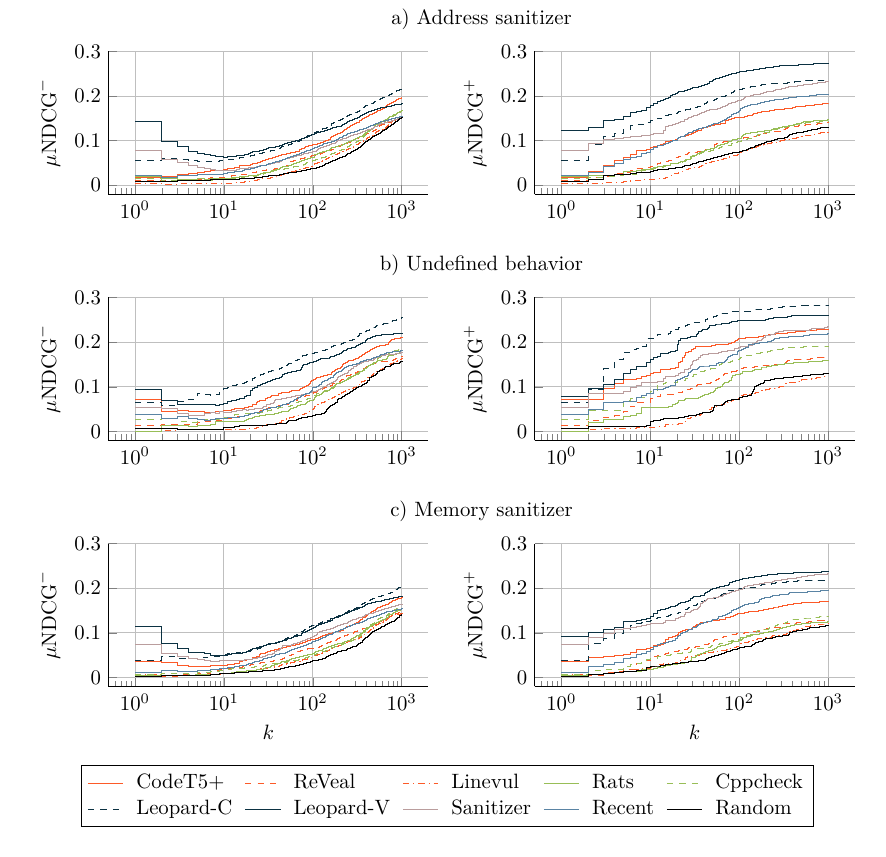}  
\end{center}
% \end{figure*}

\section{Retrieval Performance by Crash Types}
\label{sec:appx-break-down-crashtypes}

\vspace{2em}

\begin{center}
    \includegraphics[width=1.00\linewidth]{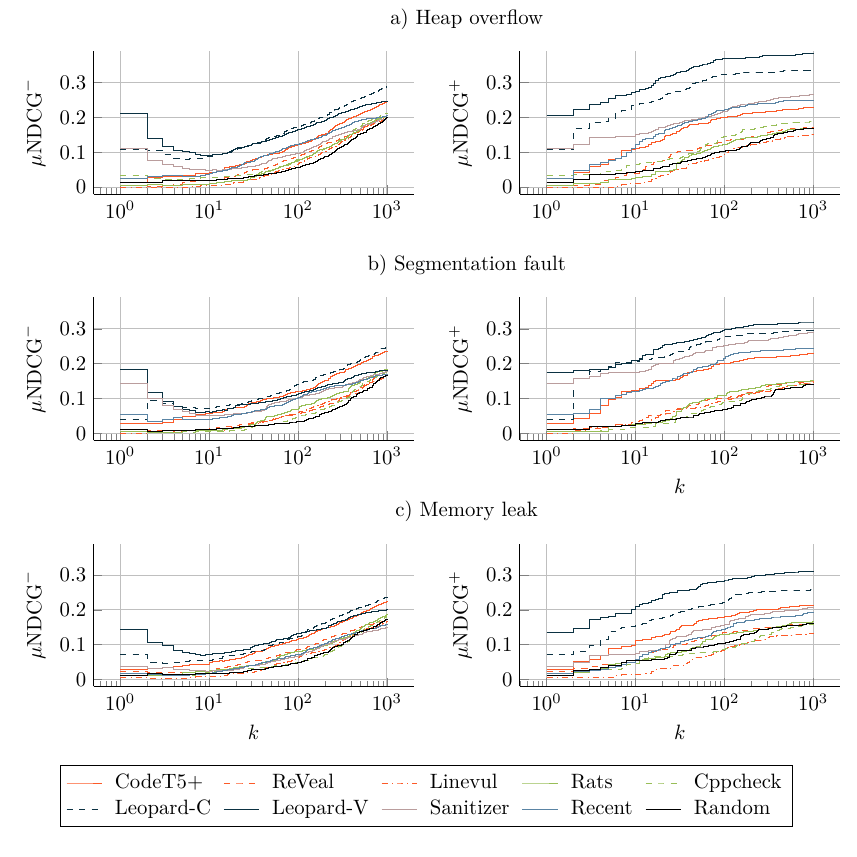} 
\end{center}

\end{document}